\documentclass[runningheads]{llncs}
\usepackage{graphicx}
\usepackage{cite}
\usepackage{amsmath,amssymb,amsfonts}
\usepackage{algorithmic}
\usepackage{graphicx}
\usepackage{textcomp}
\usepackage{xcolor}
\usepackage{wrapfig}
\usepackage{subfig}
\begin{document}
\title{Dynamic pricing and discounts by means of interactive presentation systems in stationary point of sales\thanks{This research was partially founded by Wroclaw University of Science and Technology and Poznan University of Economics and Business statutory funds, and BiWISS grant agreement no INNOTECHK3/IN3/56/225874/NCBR/15 of European Union Regional Development Fund Third Programme Innotech, IN-TECH.}}
\titlerunning{Dynamic pricing and discounts}
% If the paper title is too long for the running head, you can set
% an abbreviated paper title here
%
\author{Marcin Lewicki\inst{1} \orcidID{0000-0002-3526-487X} \and
Tomasz Kajdanowicz \inst{2} \orcidID{0000-0002-8417-1012} \and
Piotr Bródka \inst{2}  \orcidID{0000-0002-6474-0089} \and
Janusz Sobecki \inst{2}  \orcidID{0000-0001-7444-2627}}
\authorrunning{M. Lewicki et al.}
% First names are abbreviated in the running head.
% If there are more than two authors, 'et al.' is used.
%
\institute{Poznań University of Business and Economics, Poznań, Poland 
\email{marcin.lewicki@ue.poznan.pl} \and
Wrocław University of Science and Technology, Wrocław, Poland}
\maketitle % typeset the header of the contribution
\begin{abstract}
The main purpose of this article was to create a model and simulate the profitability conditions of an interactive presentation system (IPS) with the recommender system (RS) used in the kiosk. 90 million simulations have been run in Python with SymPy to address the problem of discount recommendation offered to the clients according to their usage of the IPS. 

\keywords{consumer behaviour \and recommendation system \and discount \and price discrimination}
\end{abstract}

\section{Introduction}

Nowadays, with the constantly increasing competition from the Internet stores, convincing a consumer to buy something from a stationary point of sale (PoS) is becoming a lot harder. However, with the help of new technologies: tablet recommendation algorithms, automatic customer profiling, and machine learning, even a stationary PoS can present an offer which is tailored to specific consumer needs. When these Pos form a network, then we also could profit from the application of Internet technologies to exchange online specific marketing information. Another advantage of modern technology is the possibility to build models which allow, using the knowledge about a consumer behaviour, to recommend the seller appropriate marketing actions in a real environment. Using these models, we can estimate the potential value of a new technology or an approach without conducting expensive trials, which is especially important during the earliest stages of idea development.

Recommender systems (RS) are designed to deliver customised information for very differentiated users in many different domains~\cite{montaner2003}. In~\cite{zhang2018explainable} the following areas of explaiable RS have been distinguished: e-commerce, POI, social, and multimedia recommendations. It is worth to note that the first online recommender system (ORS) was Tapestry~\cite{goldberg1992}, which was designed to filter emails. In the recommender systems taxonomy proposed by Montaner et. al.~\cite{montaner2003} we can consider the following dimensions: profile generation and maintenance, and profile exploitation. The first dimension contains the following elements: user profile representation, initial profile generation, profile learning technique, and relevance feedback. The second dimension considers information filtering method, user profile–item matching technique, user profile matching technique, and profile adaptation technique. One of these elements is of special importance, i.e., the filtering method. We can distinguish three basic filtering methods: demographic filtering (DF), content-based filtering (CBF), and collaborative filtering (CF). Nowadays also a hybrid approach, using two or more basic methods is quite often used. In the e-commerce applications of ORS that are based on customers’ online history, the most common filtering methods are CBF and CF~\cite{jiang2015}. Besides research on RS development, we can also find examples of studies concerning the impact of RS on the customer’s purchase decision. For example, in a recent study~\cite{ilicheva2015discounts} the comparative analysis of pricing promotions that formulates the willingness to buy in fashion e-shops from two distinct markets in Russia and Sweden is given. In other study~\cite{jiang2015} application of price promotion together with product recommendation should be considered for optimal profits. It tries to find the answers to the following questions: (i) how to determine the recommended product and how to price the discounted product given the specific customers shopping attitudes, (ii) what is the impact of recommendation on the discount, (iii) how changes in the cost of the product, recommendation accuracy and complementarity influence the e-tailer’s profit.
In previous years, research concentrated on the application of a shopbot that finds savings for a customer on product promotion~\cite{garfinkel2006} as well as a recommendation system that enhances the profit of e-tailers~\cite{wang2009}.

The main research objective of this article is to use the present state of knowledge about consumer behaviour in the context of dynamic pricing and discounts (including primarily their influence on a consumer intention to buy) to build a model which will allow to understand the possible impact of the interactive presentation system (IPS) developed within the Polish National Centre for Research and Development grant entitled “Business Intelligence Tools for Virtual Network of Points of Sale” (BIWiSS) on the increase of profits in the stationary point of sales where the system is installed.

The article is structured as follows. In Sec.~\ref{interactive} a general concept of interacting presentation systems is illustrated. Next, the role of discounts in retail is addressed (Sec.~\ref{discount}) and problems of price discrimination (Sec.~\ref{price}) The experimental environment is presented (Sec.~\ref{setup}). In Sec.~\ref{results} the key findings of the experiment are described and discussed. The paper is concluded in Sec.~\ref{summary}.

\section{Interactive presentation system}
\label{interactive}

Presently, we are surrounded by modern digital interactive media almost everywhere, especially in public places. There are many types of interactive media applications and configurations, which are used in many application areas such as advertisement, commerce, entertainment, and education. These applications are using many different technologies such as touch screens, kinetic interfaces, interactive walls and floors, Augmented Reality, Virtual Reality, and many others~\cite{anisiewicz2015configuration}.
One of the forms of modern interactive media is digital signage that’s main application is advertisement targeted to large audiences at public venues~\cite{chen2009interacting}. Many authors report that they deliver an effective form of media and offer many improvements in consumers perceptions of the presented information. The Digital Signage Installations may be characterized by the following factors~\cite{anisiewicz2015configuration}: optimal communicativeness, utilitarianism of solutions, synergic links of physical form with software, hybrid character of the structure, localisation in architectonic spaces, the phenomenon of attracting spectators, and ambient character.
The hybrid character of structure from the above-mentioned property list defines the mixing of building materials and the information technology that is placed in the real space. A sample Digital Signage installed in a shopping mall produces the revenue stream by charging advertisers improving ‘atmosphere’ and image by delivering interesting, informative and entertaining content at the same time~\cite{newman2010shoppers}. Application of Digital Signage systems has many advantages over traditional presentation media: possibility to present dynamic multimedia content, scalability, flexibility, interactivity of presentation and reduction of costs in a long term usage.
Taking all this into account, the multimedia shopping information system has been developed.
%, see Figure~\ref{fig:example_system}. 
The main purpose of this system was to deliver and monitor marketing information to the customers of the network of expert points of sale of GSM industry products that includes sale and service of mobile phones and their accessories.

The system is built out of two applications: content management and presentation. The main functionality of the first application consists of the following elements: presentation project design, presentation project management, presentation system management that consists of multiple consoles, usage, data gathering, rapport generation, and system administration and authorization. The second application is quite simple, and its functionality consists of two main elements: presentation of the marketing offer on touch screens and the tracking of the particular offer selection.
The system also enables real-time monitoring of the customer usage of the presentation console by the salesmen. They can monitor which offers the customer paid attention to and how long he or she were looking at the information on the particular products. The system also enables the ex-post analysis of viewing data in the whole network of presentation consoles considering not only the localization of the console but also the time of the event. These data may also be analysed together with the sales data coming from the cash register.

\section{Discounts in retail}
\label{discount}
The role of discounts in retail has been widely discussed in the literature over the years. Undoubtedly, one of the main reasons for this interest is the simple fact that despite the possibility of using numerous promotion tools, discounts remain the most frequently used one. It is no exaggeration to say that modern consumers live in the world of permanent discounts. An obvious problem is the answer to the question about the impact of discounts on sales. Studies have shown that price reductions affect consumer’s behaviour and are often used by companies to simply boost their sales~\cite{inman1993retailer}, nevertheless, it should be emphasised that the answer to the above issue is relatively complex, becoming a major factor stimulating further discussion in the area. It is also one of the main reasons why the extent of the individual publications on the topic is highly diversified. 

Gupta and Cooper~\cite{gupta1992discounting}, based on previous studies in the 80s, examined the consumer’s response to the retailer’s price promotions, showing that consumers actually discount price discounts (i.e. consumer’s perceptions of discounts are typically less than the advertised discount) depending on the discount level, store image, and whether the advertised product is a name brand or a store brand. Key findings from their research included the following: (i) consumer’s intentions to buy do not change unless the promotional discount is above a threshold level, (ii) the threshold level for a name brand is lower than that for a store brand, (iii) there is a promotion saturation point above which the effect of discounts on changes in consumer’s intention to buy is minimal.

A few years later, in 1998, Chen, Monroe and Lou~\cite{chen1998effects} attempted to determine the significance of framing price promotion and how does it affect the consumer’s perception and intentions to buy. The authors created an original framework providing recommendations for framing price discounts, depending on the relative price level of the promoted product and the size of a price decrease. It was suggested that: \textbf{(i)} for relatively low-price products and small price decrease, emphasis should be on providing consumers the absolute amount of price reduction, \textbf{(ii)} for relatively large price reductions on low-price items, the emphasis should be on the relative savings, \textbf{(iii)} for relatively high-price items, the emphasis should be on providing absolute savings, and if the price reduction is large, the relative amount of reduction as well.

Interesting conclusions were also presented by Alford and Biswas~\cite{alford2002effects}. They studies focused on a discussion about the effects of discount level, price consciousness and sale proneness on consumer’s price perception and behavioural intention. They found that a higher discount level is significantly reducing search intentions and increases perceptions of the value of the offer and intention to buy (which was consistent with some of the previous findings from the authors). Moreover, the results from their research showed that consumers with higher levels of price consciousness perceive a high level of benefits of an additional search regardless of the discount level or their level of sale proneness. Whereas this finding could be expected (the more consumers focus on paying a low price, the more they should search for the best discount), surprisingly it was noted that there was no effect of the price consciousness variable on buying intention. The authors suggested that it is possible that consumer judgement of value and buying intention could be addressed only after the most direct judgement, i.e., whether consumers can obtain the product for a lower price at another store. Finally, Alford and Biswas suggested that a dual strategy of offering a discount to affect value perceptions and buying intention and using low price guarantees to affect search intention might be an effective strategy for retail merchants. 

In 2005, Drozdenko~\cite{drozdenko2005risk} addressed important questions within the subject of discount levels, i.e., what is the risk and maximum acceptable discount levels. The study of 453 consumers who were asked to choose their own optimal discount levels (from 0-80\%) for eight categories across two distribution channels (physical stores and on-line merchants) revealed strong consumer perceptions about discount risks and the trade-offs consumers make between risk and financial benefits across different product categories (regardless whether it is and on-line or off-line sale) which undoubtedly could help retailers in setting optimal discount levels. It was found that setting the highest possible level of reduction does not mean that a retailer will sell more products/services. As it turned out, only 13\% of respondents selected the 80\% discount level for each product and each channel, despite seeing the exact price they would pay at each level. One of the reasons for such results is the risk perceived by consumers. 88\% of them attributed at least one cause for the deepest discounts. Most frequently cited were concerns about quality problems, damaged goods, or stolen goods. Additionally, it was also found that consumers opted for lower discount levels from the on-line merchant than from the physical store. Finally, there was a wide divergence by product category, with consumers selecting smaller discounts on tires and cereals and the deepest discounts on shirts. Thus, it is impossible to set universal rules about optimal discount levels that will have the same effect on the consumer’s intention to buy or a perceived risk regardless of the product or the market. 

A brief overview of the recent publications on the topic shows some of the directions of nowadays research. Regardless of whether the article is focused on: the analysis of the impact of price promotion strategies as a motivating factor for the manufacturer and his sales performance~\cite{cui2016analyzing}, the use of fake discounts~\cite{ngwe2018fake},the profitability of stacked discounts~\cite{ertekin2019profitability} or the use of gambled price discounts to enhance subscription-based e-commerce services~\cite{tan2021enhancing}, it is quite obvious that a room for a discussion on the subject is still present. Moreover, taking into account all possible factors that could influence the effectiveness of discounts, it is really hard to imagine a situation in which the existing cognitive gap within this topic will be ever closed.

\section{Price discrimination}
\label{price}

The phenomenon of selling the same commodity at different prices to different consumers, which is considered the optimal method of pricing (from the economic point of view~\cite{mcafee2008price, acquisti2005conditioning,varian1989price}) has been well described within the economic literature over the years. Some of the most significant studies in the area came from 80’s (e.g.~\cite{phlips1983economics,varian1989price,tirole1988theory} and did serve as a foundation for further research in the area. There are three different types of price discrimination (based on Pigou classification from 1920~\cite{miller2014we}) i.e.:
\textbf{(i)} First-degree, or perfect price discrimination involves the seller charging a different price for each unit of the good in such a way that the price charged for each unit is equal to the maximum willingness to pay for that unit, 
\textbf{(ii)} Second-degree price discrimination, or nonlinear pricing, occurs when prices differ depending on the number of units of the good bought, but not across consumers. That is, each consumer faces the same price schedule, but the schedule involves different prices for different amounts of goods purchased,
\textbf{(iii)} Third-degree price discrimination means that different purchasers are charged different prices, but each purchaser pays a constant amount for each unit of the good bought.

Pricing strategy depends on the seller’s ability to take advantage of the information exchanged during the commercial relationship. A first-degree price discrimination strategy requires that the firm is able to uniquely identify each consumer. It also requires a lot of information about the consumer’s tastes and the highest willingness to pay to tailor a price to an individual consumer. The second one requires general information about the dispersion of price-sensitivities among consumers to construct an efficient menu of options. The third one requires that the seller can identify at least whether the consumer has the relevant group trait that is used for discrimination~\cite{miller2014we}.
Looking at the problem of price discrimination from the perspective of the past few years it is hard not to mention the role of the new information technologies. Despite the fact that shoppers gained access to new price comparison tools, the vision of fully informed and empowered consumers did not materialise. It is mostly due to the fact that new information technologies also continue to create new opportunities for tailoring prices to individual consumers – one of them being dynamic pricing~\cite{miller2014we}. Undoubtedly the extensive information collected online from consumers stimulates price discrimination, moreover, the information itself often leads to price discrimination~\cite{mikians2012detecting}. Some recent studies further explore personalised pricing~\cite{borgesius2017online} or the subject of behavioural constraints on price discrimination~\cite{leibbrandt2020behavioral}and therefore are leading to a conclusion that price discrimination is a rather fixed point in modern marketing strategies.

\section{Experimental setup}
\label{setup}

The simulation process is representing the context of a purchase process in a mall kiosk. Mall kiosk is a small retail outlet located in the aisle of a shopping mall. In essence, a customer of a kiosk can browse the on-site visible goods and approach to the counter to purchase them. 

We can assume the most straightforward scenario of the possible customer’s actions in the purchase process. It can be characterized by the following steps:
\begin{enumerate}
\item Customer enters the kiosk shopping area.
\item It may directly specify its buying needs to the salesmen, who prepares the goods and serves with the payment process, which ends up the purchase. 
\item However, the customer can approach the interactive display prior to the counter and browse through kiosk offer.
\item While checking the available goods, the customer has the possibility to browse the list of categories of goods, listing items in the category, as well as getting to details on particular products. 
\item The interactive display registers all actions, as well as the time, spend on each screen with product details. On the other side of the system, the salesmen get presented the good (list of goods) that gathered the most attention of the customer on display, measured in time units.
\item While leaving the display the customer either goes to the register or is offered a special offer by the salesmen, according to the preference revealed in the interactive display. The offer has the form of discount in original price. 
\item As an effect customer can buy or not the product taking into account the offered discount.
\end{enumerate}
 
To represent the purchase process with proper expression of customers needs and profiles it has been decided to build a model that consists of seven following parameters: \textbf{(i)} indicator whether a customer approaching the shopping area uses an interactive display, \textbf{(ii)} a category of a product that was the most observed by the customer at interactive display, \textbf{(iii)} price of the product that the customer watched the most at interactive display and probably is willing to buy, \textbf{(iv)} customer's initial purchase intention, \textbf{(v)} a discount given to the customer for the discovered product of interest, \textbf{(vi)} linearised form of discount related purchase intention increase law, based on~\cite{gupta1992discounting}, \textbf{(vii)} kiosk's overhead (called here margin) included in the price.

\begin{figure}[t]
\centering
\subfloat{\includegraphics[width=.45\textwidth]{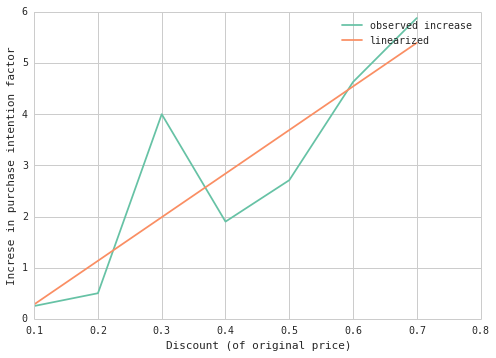}}\hfill
\subfloat{\includegraphics[width=.45\textwidth]{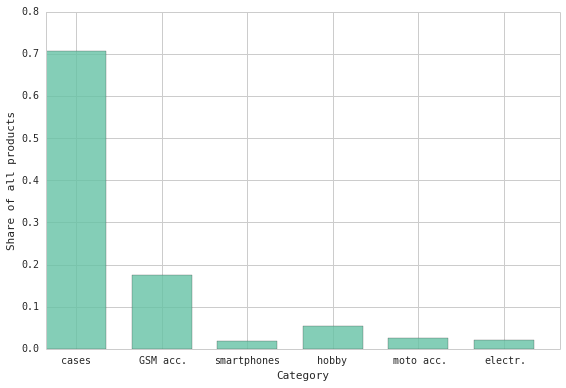}}\hfill
\caption{Purchase intention increase as a function of discount(a) and share of particular product categories (b).}
\label{fig:categories_purchase_intention_increase}
\end{figure}
\label{results}

In general, the model can represent a population of customers by assuming appropriate distributions of the parameters. In the model the customer that receives a discount increases his initial purchase intention, and therefore, rises the chance to buy the product. The outcome of the parametrized model is a quantification of the overheads (sum of margins).

The idea to parametrize the model for simulation purposes is to define the probability whether a customer approaching the shopping area uses an interactive display by a binomial distribution sampled for p=(0.1,0.7) with 0.02 step. It means that by means of distribution it is tested the behaviour of customers from 10\% of them using the interactive display to 70\% of the population with simulation step equal to 2\%. The category of product that was most interesting for the customer using an interactive display is modeled as a random choice of one out of six product categories. It is the one which the customer spends the most time in while browsing the interactive display. The number of categories is inspired by the real-world kiosk case that sells telecommunication products in multiple locations in Poland (\url{www.teletorium.pl}). The categories are 'cases and protectors', 'GSM accessories', 'smartphones and tablets', 'hobby \& sport', 'moto accessories', 'electronics'. Random choice of category is weighted by a share of a number of products in particular category to the sum of all products in all categories. It was assumed that popularity of particular category is related to the number of products in it, fig.~\ref{fig:categories_purchase_intention_increase} (a). The price is sampled from the previously estimated normal distribution of prices of products within each category - similarly based on mentioned real-world kiosk. It means that each category was checked regarding price distribution and Gaussian approximation of that distribution was proposed. The parameters of the distribution were the following: 'cases and protectors': $N(29,8)$, 'GSM accessories': $N(35,8)$, 'smartphones and tablets': $N(700,200)$, 'hobby \& sport': $N(45,10)$, 'moto accessories': $N(80,21)$, 'electronics': $N(50,13)$. Customer's initial purchase intention was modeled by a parameter interpreted as the probability the product will be bought and was taken from the range (0.1, 0.7) with step 0.02. The discount parameter that is an offer of the salesmen were taken from the range (0.1, 0.7) with step 0.02. This same was simulated discounts from 10\% to 70\% of original prices. The linearised form of purchase intention increase thanks to discount is shown in fig.~\ref{fig:categories_purchase_intention_increase} (b) and expressed in $\mathbf{PII=8.52 \times D - 0.57}$, where $PII$ denotes purchase intention increase, $D$ is a discount. The simulation was performed for three distinct values of margin (overhead) $\{0.3, 0.4, 0.5\}$, based on the average real-world kiosk case. The aim of the simulation was to test the proposed model for all permutations of the parameters' values. Each simulation assumes that the kiosk is visited by 1000 customers. They generate some turnover and depending on the configuration of the model (especially the discount and the margin) generate some profit or loss.

\section{Results and discussion}
The simulation had to consider 89,373 permutations of parameters, for each permutation of the model parameters, it performed a behaviour simulation of 1000 customers. The simulation was performed by a programme written in Python language with the usage of SymPy module (\url{www.sympy.org}). Through careful consideration of the results, it was desired to acquire the profitability conditions of interactive display utilization in the kiosk. Therefore, the results contained roughly 90 millions of records. The presentation of them is split into the analysis of loss in margin (overhead), analysis of customers' number as well as a generalized profitability statement. End to end, the last part of this section shows when it worth to deploy the proposed interactive display. 

\begin{figure}[t]
 \centering
 \raisebox{35pt}{\parbox[b]{.1\textwidth}{Avg. margin 0.3}}%
 \subfloat[PI 0.1]{\includegraphics[width=.28\textwidth]{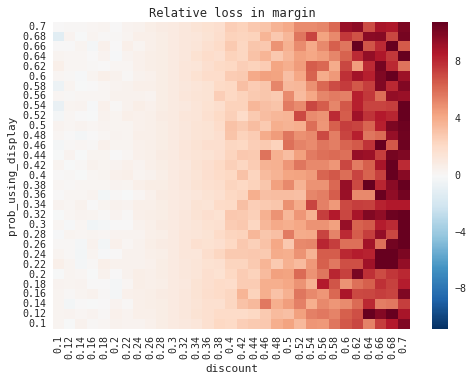}}\hfill
 \subfloat[PI 0.36]{\includegraphics[width=.28\textwidth]{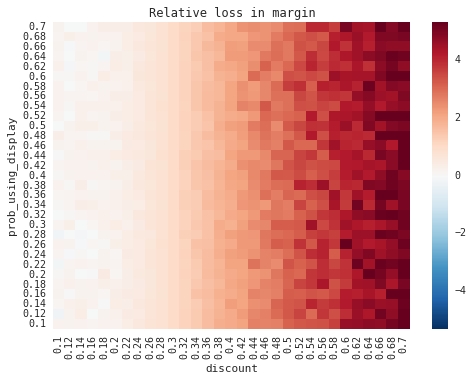}}\hfill
 \subfloat[PI 0.7]{\includegraphics[width=.28\textwidth]{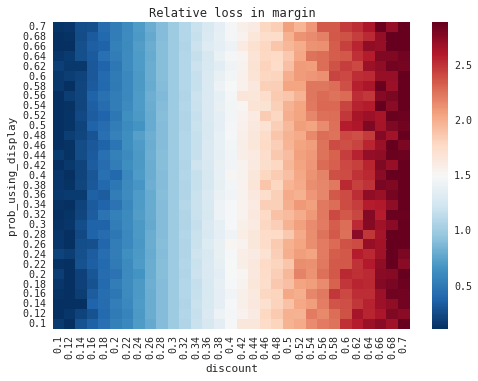}}\par
 \raisebox{35pt}{\parbox[b]{.1\textwidth}{Avg. margin 0.4}}%
 \subfloat[PI 0.1]{\includegraphics[width=.28\textwidth]{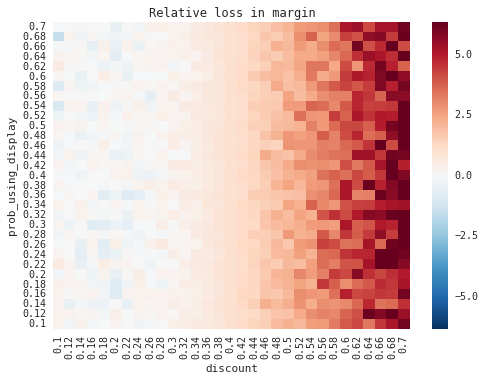}}\hfill
 \subfloat[PI 0.36]{\includegraphics[width=.28\textwidth]{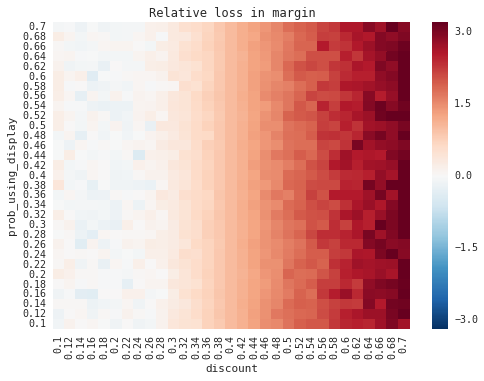}}\hfill
 \subfloat[PI 0.7]{\includegraphics[width=.28\textwidth]{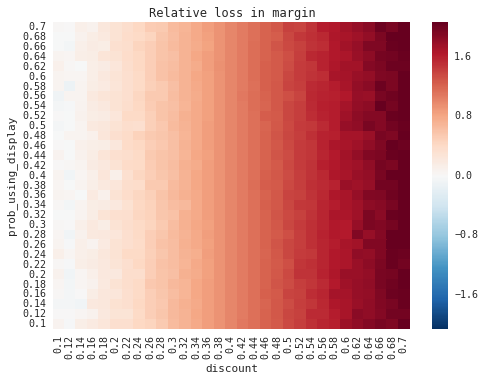}}\par
 \raisebox{35pt}{\parbox[b]{.1\textwidth}{Avg. margin 0.5}}%
 \subfloat[PI 0.1]{\includegraphics[width=.28\textwidth]{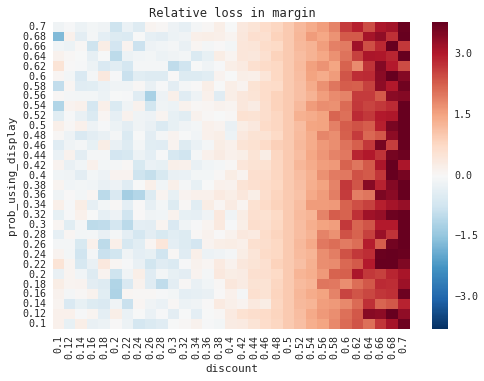}}\hfill
 \subfloat[PI 0.36]{\includegraphics[width=.28\textwidth]{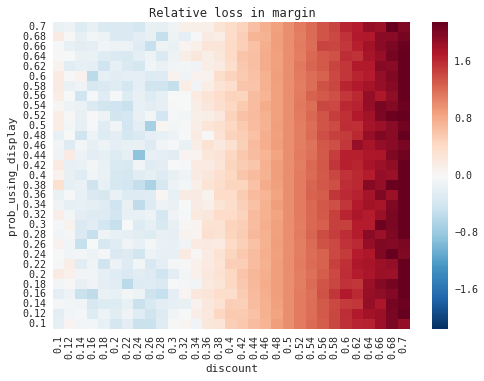}}\hfill
 \subfloat[PI 0.7]{\includegraphics[width=.28\textwidth]{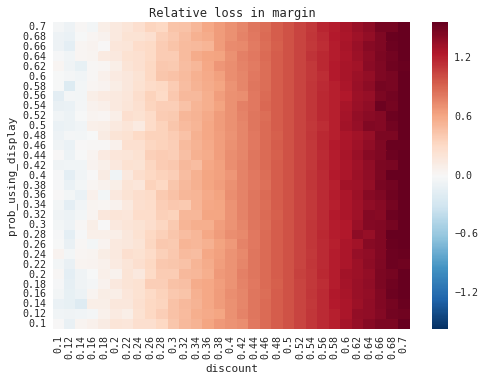}}
 \caption{The results of relative loss in margin for distinct permutations of parameters. (PI - Purchase intention)}
 \label{fig:margin_loss}
\end{figure}

\subsection{Loss in margin}
The results presented in fig.~\ref{fig:margin_loss} shows the relative loss that was generated in the kiosk for distinct combinations of parameters. In the fig.~\ref{fig:margin_loss} , we can observe the relative loss (or in some cases the profit) between two scenarios: the sum of overheads in a situation when the interactive display is not present in the kiosk and the sum of overheads when there is an interactive display available and used by some proportion of customers. By performing the second scenario, we must grant a discount to those customers who would potentially make a purchase but had too low purchase intention and while discounting products we raise it. In other words, we decrease the overhead (margin) by giving a discount to the customer that would not buy the product we have identified as interesting for that customer by means of the interactive display.

One straightforward observation is that regardless of what proportion of customers will be using, the interactive display loss in margin (or profit) depends only on the initial purchase intention of the customers and the granted discount. We can also observe that the higher the initial margin, the easier situation to obtain profit. Interestingly, the higher initial purchase intentions, the harder situation to obtain profit. This reveals a common-sense mechanism; you do not have to convince already convinced to purchase. We can also observe that the higher the discount, the more loss is generated. However, while discussing the findings, we must take into account that the discounting mechanism is rarely used for profit purpose~\cite{levy2004emerging}. It should rather gather more customers (more purchases) at a reasonable cost. Therefore, the results of the analysis of customer number should be considered in parallel. 

To give an example of an interpretation of the results, we can state that in a situation when customers have very low initial purchase intention (0.1) having a high margin (0.5) giving a discount up to 30\% may result in a three times higher sum of margin (overhead). On the other hand, giving the high discount (70\%) with lower margin (0.3), we may loss eight times that much as it would be got without giving discounts (without interactive display). 

\begin{figure} [t]
\centering
\subfloat{\includegraphics[width=.47\textwidth]{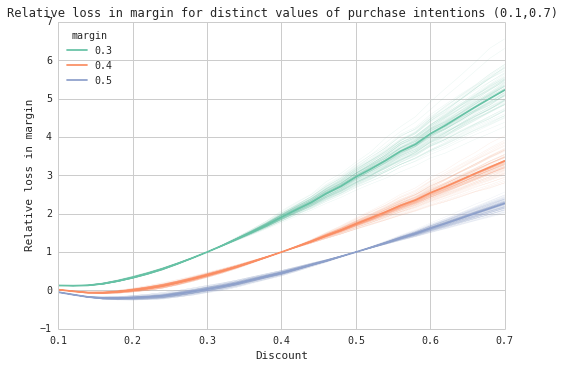}}\hfill
\subfloat{\includegraphics[width=.42\textwidth]{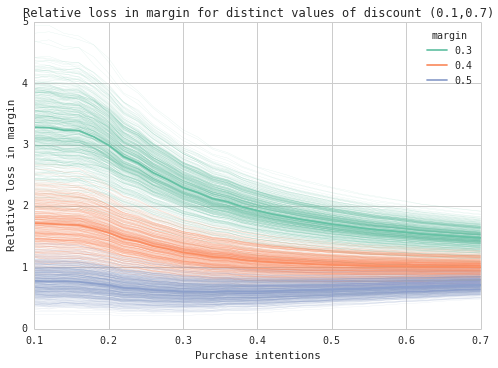}} 
\caption{Averaged results of relative loss in margin.}
\label{fig:agg_margin_loss}
\end{figure}

When we average the results of relative loss in margin (in overhead) for distinct values of initial purchase intention or for distinct values of discounts, fig.~\ref{fig:agg_margin_loss}, we can see that only circumstances for reaching the profit regardless of initial purchase intention are achieved while having a discount lower than 20\% and margin at least at 0.4 level.

\begin{figure}[t]
 \centering
 \subfloat[Purchase intention 0.1]{\includegraphics[width=.33\textwidth]{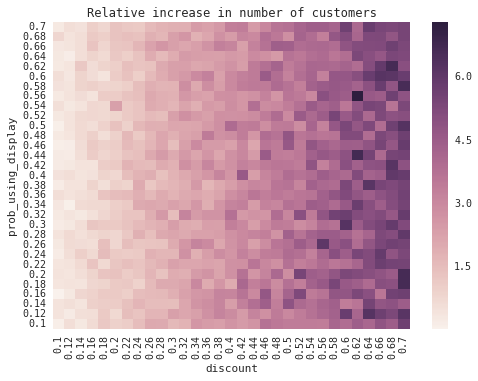}}\hfill
 \subfloat[Purchase intention 0.36]{\includegraphics[width=.33\textwidth]{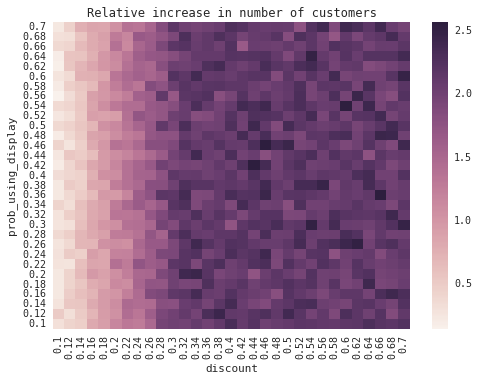}}\hfill
 \subfloat[Purchase intention 0.7]{\includegraphics[width=.33\textwidth]{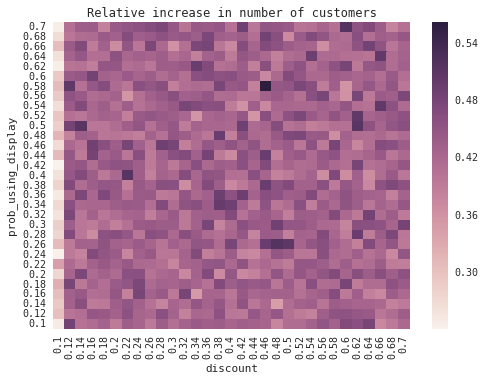}}\par
 \caption{The results of relative increase in number of buying customers for distinct permutations of parameters.}
 \label{fig:cust_increase}
\end{figure}

\subsection{Number of customers}
Similarly to the previous results, the ones presented in fig.~\ref{fig:cust_increase} shows the relative measure comparing two scenarios: without discounting (without interactive display) and with discounts (with interactive display). The fig.~\ref{fig:cust_increase} represents the relative increase of the number of purchases (number of buying customers). In general, by giving discounts, we raise the initial purchase intention and therefore increase the number of positive transactions. Generally speaking, regardless of all considered parameters, the relative increase in some customers is always positive. Moreover, the higher discounts, the higher number of customers (up to six times more than without discounts). Similarly, the lower initial purchase intentions, the higher increase of the number of customers. 

\begin{figure}[b]
\centering
\subfloat{\includegraphics[width=.485\textwidth]{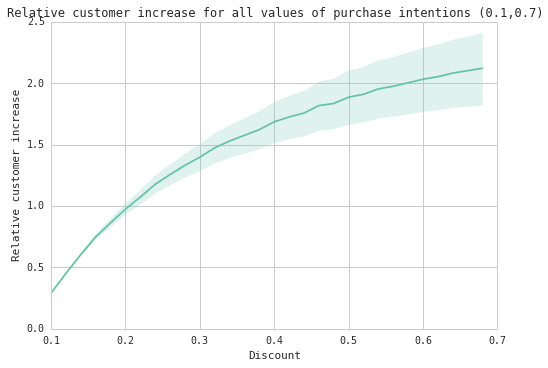}}\hfill
\subfloat{\includegraphics[width=.445\textwidth]{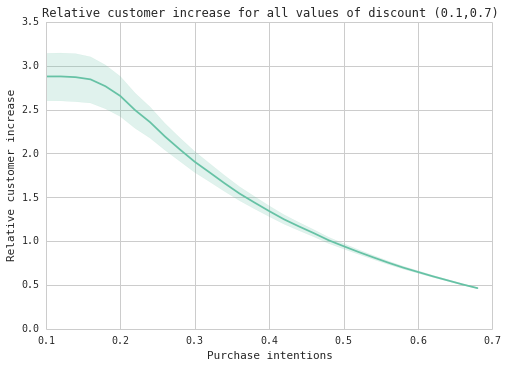}}
\caption{Aggregated results of relative increase in number of buying customers.}
\label{fig:agg_cust_increase}
\end{figure}

The phenomenon where giving the higher discount we achieve a higher number of purchases is observed only for some initial purchase intention threshold (around 0.4). In other words, increasing the discounts while having high initial purchase intention does not change anything, fig.~\ref{fig:cust_increase} (c).

When we average the results of a relative increase in buying customers for distinct values of initial purchase intention or for distinct values of discounts, fig.~\ref{fig:agg_cust_increase}, we see that regardless of initial purchase intention increasing the discount rises the number of purchases in a logarithmic fashion and the higher the initial purchase intention the near-linearly smaller the relative increase of buying customers.

\subsection{Generalized profitability}
\begin{figure}[t]%{l}{0.46\textwidth}[b]
\centering
\includegraphics[width=.45\textwidth]{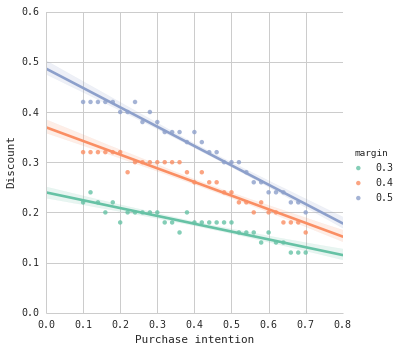}
\caption{Profitability results - area under the lines denotes profit while applying interactive display and discounts, above loss, in comparison to scenario without display and discounts.}
\label{fig:profitability}
\end{figure}

In this subsection, all results are gathered in a single fig.~\ref{fig:profitability} that can be used for assessing whether it is worth to deploy the interactive display and offer personalized discounts. As we saw in the previous subsections, the level of margin discount and initial purchase intention are important for the outcome measured in profit/loss and the number of buying customers. Fig.~\ref{fig:profitability} contains all these parameters and three straight lines in the fig. represent the equality condition - when the cost of discounts is covered by an increase in the sum of margin caused by an increase of purchases. In other words, we do not lose comparing to a scenario with no discounts and no interactive display. The area below the line denotes the situation when we observe profit and, in contrary, the loss. One should assess a kiosk's situation according to its average margin as well as the characteristics of customers - initial purchase intentions. For instance, a kiosk with an average margin on the level of 0.3 may benefit from an increase of buying customers by deploying the interactive display only with 12-24\% discounts, depending on customers' initial purchase intentions. The fig.~\ref{fig:profitability} can be used in the following way: get to know your customers' purchase intentions (e.g., count how many of the customers in your kiosk area are buying), choose an appropriate average margin, and check how much you can discount. The expected relative increase in the number of buying customers can be then read from fig.~\ref{fig:agg_cust_increase}(b).

\section{Summary}
\label{summary}
In this paper, the application and influence of an interactive presentation system has been presented. The idea has been initiated by the the project BIWiSS, where that kind of system was designed for the Polish nation-wide network of POS with smartphones and their accessories. The system consists of two main functions: content management and presentation. The first one manages the content and usage of a network of presentation systems and the second one presents the product offer of the POS. 
The main problem addressed in the paper was the recommendation of discounts offered to the clients according to their usage of the presentation system. This task is observed in any class of RS, where the recommendation fact occurs after the user is firstly willing to focus his attention on a particular offer. To examine how to deliver such a recommendation, a new model for measuring profit/loss by applying discounting scenarios in the recommendation has been introduced. To verify the model, a large number of simulations (roughly 90M) of purchases have been performed. These simulations revealed that the encountered outcome does not depend on the share of customer's using interactive display and the important parameters are customers' initial purchase intention and the level of discount. However, thanks to the application of the proposed interactive presentation system, POS can increase the number of customers by a factor of three, which is a significant outcome in building a customer base. The greater number of clients definitely builds up the potential for increasing the brand recognition and the global income of the POS network. The simulation of the proposed model has also allowed to show the profitability circumstances regarding discount and initial purchase intentions of customers to keep profitability at the same level as in the scenario without application of the system. It should be noted that performed experiments have some limitations. The behaviour of customers can be affected by a number of factors that have been not addressed within this article. Nevertheless, it is also believed that the results presented in the article are a good basis for further testing of the system in a real environment, including Social Recommender Systems. %\cite{kazienko2011multidimensional}.
\section*{Cite this paper as}
Lewicki M., Kajdanowicz T., Bródka P., Sobecki J. (2021) Dynamic Pricing and Discounts by Means of Interactive Presentation Systems in Stationary Point of Sales. 
In: Paszynski M., Kranzlmüller D., Krzhizhanovskaya V.V., Dongarra J.J., Sloot P.M. (eds) Computational Science – ICCS 2021. ICCS 2021. Lecture Notes in Computer Science, vol 12745. Springer, Cham. 
\url{https://doi.org/10.1007/978-3-030-77970-2\_46}
%
% ---- Bibliography ----
%
% BibTeX users should specify bibliography style 'splncs04'.
% References will then be sorted and formatted in the correct style.
%
\bibliographystyle{splncs04}
\bibliography{sample.bib}

\begin{thebibliography}{10}
\providecommand{\url}[1]{\texttt{#1}}
\providecommand{\urlprefix}{URL }
\providecommand{\doi}[1]{https://doi.org/#1}

\bibitem{acquisti2005conditioning}
Acquisti, A., Varian, H.R.: Conditioning prices on purchase history. Marketing
  Science  \textbf{24}(3),  367--381 (2005)

\bibitem{alford2002effects}
Alford, B.L., Biswas, A.: The effects of discount level, price consciousness
  and sale proneness on consumers' price perception and behavioral intention.
  Journal of Business research  \textbf{55}(9),  775--783 (2002)

\bibitem{anisiewicz2015configuration}
Anisiewicz, J., Jakubicki, B., Sobecki, J., Wantula, Z.: Configuration of
  complex interactive environments. In: New Research in Multimedia and Internet
  Systems, pp. 239--249. Springer (2015)

\bibitem{borgesius2017online}
Borgesius, F.Z., Poort, J.: Online price discrimination and eu data privacy
  law. Journal of consumer policy  \textbf{40}(3),  347--366 (2017)

\bibitem{chen2009interacting}
Chen, Q., Malric, F., Zhang, Y., Abid, M., Cordeiro, A., Petriu, E.M.,
  Georganas, N.D.: Interacting with digital signage using hand gestures. In:
  International Conference Image Analysis and Recognition. pp. 347--358.
  Springer (2009)

\bibitem{chen1998effects}
Chen, S.F.S., Monroe, K.B., Lou, Y.C.: The effects of framing price promotion
  messages on consumers' perceptions and purchase intentions. Journal of
  retailing  \textbf{74}(3),  353--372 (1998)

\bibitem{cui2016analyzing}
Cui, B., Yang, K., Chou, T.: Analyzing the impact of price promotion strategies
  on manufacturer sales performance. Journal of Service Science and Management
  \textbf{9}(02), ~182 (2016)

\bibitem{drozdenko2005risk}
Drozdenko, R., Jensen, M.: Risk and maximum acceptable discount levels. Journal
  of Product \& Brand Management  \textbf{14}(4),  264--270 (2005)

\bibitem{ertekin2019profitability}
Ertekin, N., Shulman, J.D., Chen, H.: On the profitability of stacked
  discounts: Identifying revenue and cost effects of discount framing.
  Marketing Science  \textbf{38}(2),  317--342 (2019)

\bibitem{garfinkel2006}
Garfinkel, R., Gopal, R., Tripathi, A., Yin, F.: Design of a shopbot and
  recommender system for bundle purchases. Decision Support Systems
  \textbf{42}(3),  1974--1986 (2006)

\bibitem{goldberg1992}
Goldberg, D., Nichols, D., Oki, B.M., Terry, D.: Using collaborative filtering
  to weave an information tapestry. Communications of the ACM  \textbf{35}(12),
   61--70 (1992)

\bibitem{gupta1992discounting}
Gupta, S., Cooper, L.G.: The discounting of discounts and promotion thresholds.
  Journal of consumer research  \textbf{19}(3),  401--411 (1992)

\bibitem{ilicheva2015discounts}
Ilicheva, E.: Discounts as a marketing tool for attraction and retention of
  customers in e-commerce through the example of a comparative analysis of the
  specificity of fashion e-shops in russia and sweden (2015)

\bibitem{inman1993retailer}
Inman, J.J., McAlister, L.: A retailer promotion policy model considering
  promotion signal sensitivity. Marketing Science  \textbf{12}(4),  339--356
  (1993)

\bibitem{jiang2015}
Jiang, Y., Shang, J., Liu, Y., May, J.: Redesigning promotion strategy for
  e-commerce competitiveness through pricing and recommendation. International
  Journal of Production Economics  \textbf{167},  257--270 (2015)

\bibitem{leibbrandt2020behavioral}
Leibbrandt, A.: Behavioral constraints on price discrimination: Experimental
  evidence on pricing and customer antagonism. European Economic Review
  \textbf{121},  103303 (2020)

\bibitem{levy2004emerging}
Levy, M., Grewal, D., Kopalle, P.K., Hess, J.D.: Emerging trends in retail
  pricing practice: implications for research. Journal of Retailing
  \textbf{80}(3),  xiii--xxi (2004)

\bibitem{mcafee2008price}
McAfee, R.P.: Price discrimination. Issues in Competition Law and Policy
  \textbf{1},  465--484 (2008)

\bibitem{mikians2012detecting}
Mikians, J., Gyarmati, L., Erramilli, V., Laoutaris, N.: Detecting price and
  search discrimination on the internet. In: Proceedings of the 11th ACM
  Workshop on Hot Topics in Networks. pp. 79--84. acm (2012)

\bibitem{miller2014we}
Miller, A.A.: What do we worry about when we worry about price discrimination?
  the law and ethics of using personal information for pricing. Journal of
  Technology Law and Policy  \textbf{19},  43--104 (2014)

\bibitem{montaner2003}
Montaner, M., L{\'o}pez, B., De~La~Rosa, J.L.: A taxonomy of recommender agents
  on the internet. Artificial intelligence review  \textbf{19}(4),  285--330
  (2003)

\bibitem{newman2010shoppers}
Newman, A., Dennis, C., Wright, L.T., King, T., et~al.: Shoppers' experiences
  of digital signage-a cross-national qualitative study. JDCTA  \textbf{4}(7),
  50--57 (2010)

\bibitem{ngwe2018fake}
Ngwe, D.: Fake discounts drive real revenues in retail. Harvard Business School
  (2018)

\bibitem{phlips1983economics}
Phlips, L.: The economics of price discrimination. Cambridge University Press
  (1983)

\bibitem{tan2021enhancing}
Tan, W.K., Chen, B.H.: Enhancing subscription-based ecommerce services through
  gambled price discounts. Journal of Retailing and Consumer Services
  \textbf{61},  102525 (2021)

\bibitem{tirole1988theory}
Tirole, J.: The theory of industrial organization. MIT press (1988)

\bibitem{varian1989price}
Varian, H.R.: Price discrimination. Handbook of industrial organization
  \textbf{1},  597--654 (1989)

\bibitem{wang2009}
Wang, H.F., Wu, C.T.: A mathematical model for product selection strategies in
  a recommender system. Expert Systems with Applications  \textbf{36}(3),
  7299--7308 (2009)

\bibitem{zhang2018explainable}
Zhang, Y., Chen, X.: Explainable recommendation: A survey and new perspectives.
  Foundations and Trends in Information Retrieval  \textbf{14}(1),  1--101
  (2020)

\end{thebibliography}

\end{document}